\documentstyle[rmp,aps,twocolumn,epsf,amssymb]{revtex}
\newcommand{\beq}{\begin{equation}}\newcommand{\eeq}{\end{equation}}
\newcommand{\be}{\begin{equation}}\newcommand{\ee}{\end{equation}}
\newcommand{\bea}{\begin{eqnarray}}\newcommand{\eea}{\end{eqnarray}}

\begin{document}

\title{Low temperature physics at room temperature in water:
Charge inversion in chemical and biological systems}
\author
{A.Yu. Grosberg, T.T. Nguyen, and B.I. Shklovskii}
\address{Department of Physics, University of Minnesota,
116 Church Street SE, Minneapolis, Minnesota 55455}

\address{\begin{quote} {\em We review recent advances in the physics of
strongly interacting charged systems functioning in water at room
temperature.  We concentrate on the phenomena which go beyond the
framework of mean field theories, whether linear Debye-H\"uckel or
non-linear Poisson-Boltzmann.  We place major emphasis on charge
inversion - a counterintuitive phenomenon in which a strongly
charged particle, called macroion, binds so many counterions that
its net charge changes sign.  We discuss the universal theory of
charge inversion based on the idea of a strongly correlated liquid
of adsorbed counterions, similar to a Wigner crystal. This theory
has a vast array of applications, particularly in biology and
chemistry; for example, the DNA double helix in the presence of
positive multivalent ions (e.g., polycations) acquires a net
positive charge and drifts as a positive particle in electric
field.  This simplifies DNA uptake by the cell as needed for gene
therapy, because the cell membrane is negatively charged.  We
discuss also the analogies of charge inversion in other fields of
physics.}
\end{quote}}

\maketitle

\tableofcontents

\section{Introduction}\label{sec:Introduction}

Molecular biological machinery functions in water at around room
temperature.  For a physicist, this very limited temperature range
contrasts unfavorably to the richness of low temperature physics,
where one can change temperature and scan vastly different energy
scales in an orderly manner.  In this Colloquium, we review the
recently developed understanding of highly charged molecular
systems in which Coulomb interactions are so strong that we are
effectively in the realm of low temperature physics.

To be specific, imagine a problem in which one big ion, called a
{\em macroion}, is screened by much smaller but still multivalent
ions with a large charge $Ze$ each ($e$ is the proton charge); for
brevity, we call them $Z$-ions. A variety of macroions are of
importance in chemistry and biology, ranging from the charged
surface of mica or charged solid particles, to charged lipid
membranes, colloids, DNA, actin, and even to cells and viruses.
Multivalent metal ions, charged micelles, dendrimers, short or
long polyelectrolytes including DNA - to name but a few - can play
the role of the screening $Z$-ions.

\begin{figure}
\epsfxsize=8.5cm \centerline{\epsfbox
{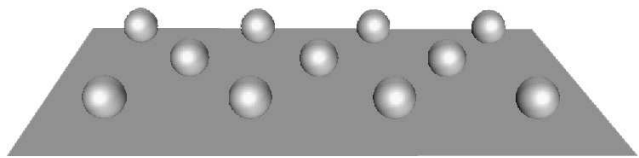}}\caption{Strongly correlated liquid (SCL) -
almost a Wigner crystal - of $Z$-ions on the oppositely charged
macroion surface.  The figure is characteristic in showing the
degree to which we are willing to ignore the microscopic details.
}\label{fig:WCpicture}
\end{figure}

The central idea of this Colloquium is that of {\em correlations}:
due to strong interactions with the macroion surface and with each
other, screening $Z$-ions do not position themselves randomly in
three-dimensional space, but form a strongly correlated liquid on
the surface of the macroion. Moreover, in terms of short range
order this liquid is reminiscent of a Wigner crystal, as the
cartoon in Fig. \ref{fig:WCpicture} depicts.  Because of its
central importance we shall use special abbreviation SCL to denote
a strongly correlated liquid of adsorbed $Z$-ions.

Depending on the system geometry and other circumstances,
correlations between screening ions may appear in many different
ways. To create some simple images in the reader's mind, it is
useful to begin with a few examples.  One example is that in Fig.
\ref{fig:WCpicture}, which may be the surface of, e.g., a latex
particle screened by some compact ions.  With a modest leap of the
imagination, we can envision the same picture to represent the
surface of a DNA double helix screened by multivalent counterions,
such as spermine with $Z=4$ \cite{Bloomfield}. Here, we imagine
DNA as a thick and long cylinder, diameter 2 nm and charge $-e$
per 1.7 nm along the cylinder, or, in other words, with a huge
negative surface charge density $-0.9 \ e / $nm$^2$.  We study
correlations between point-like $Z$-ions in Secs.
\ref{sec:small_ions} - \ref{sec:salt}.  One obvious problem with
the model Fig. \ref{fig:WCpicture} is that it ignores the
discreteness of charges, both in the macroion and in the $Z$-ions;
chemists, for instance, hardly ever accept such continuous models.
In Sec. \ref{sec:other_mechanisms}, we address the electrostatic
correlations in the systems with discrete charges.

\begin{figure}
\epsfxsize=5.5cm \centerline{\epsfbox{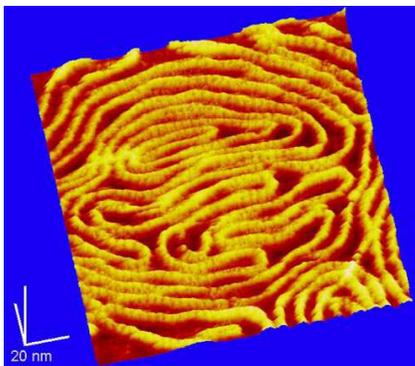}} \caption{DNA
adsorbed on a positively charged surface, as seen in an atomic
force microscopy image \protect\cite{Virginia}. }\label{fig:dna2}
\end{figure}

As another example, consider DNA molecules which screen the
positive surface charge of a colloid particle.  Obviously, DNA
chains play the role of $Z$-ions. For very short DNA pieces we are
back again to Fig. \ref{fig:WCpicture}, but longer DNA are
spaghetti-like, and in this situation, correlations mean parallel
arrangement, as we see in Fig. \ref{fig:dna2}. Theoretical
treatment of this problem in Sec. \ref{sec:rods} makes a
simplifying assumption that it may be modeled as a system of
parallel rods.

\begin{figure}
\epsfxsize=6.0cm \centerline{\epsfbox{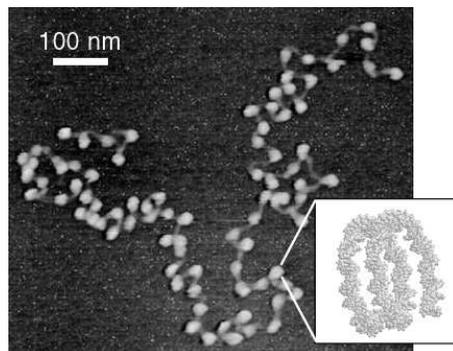}}
\caption{Electron microscopy image of 10 nm chromatin fiber
\protect\cite{Virginia2}. The beads are nucleosomes.  The
structure of a nucleosome is known to about 0.2 nm resolution
\protect\cite{XrayNucleosome}, and the inset shows how the DNA
double helix is bent in the nucleosome.  While the overall shape
of the fiber is perhaps due to the sample preparation procedure,
the array of nucleosomes on the DNA is close to periodic, similar
in this respect to the theoretical model shown in Fig.
\protect\ref{fig:necklace}. }\label{fig:chromatin1}
\end{figure}

\begin{figure}
\epsfxsize=9.0cm \centerline{\epsfbox{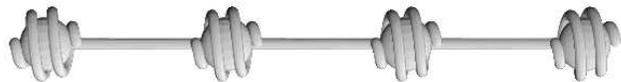}}
\caption{Self-assembled complex of a negative polyelectrolyte
molecule and many positive spheres in a necklace-like structure.
On the surface of a sphere, neighboring polyelectrolyte turns are
correlated similar to the rods in Fig. \protect\ref{fig:rods}
below. On a larger scale, charged spheres repel each other and
form one-dimensional Wigner crystal along the polyelectrolyte
molecule.}\label{fig:necklace}
\end{figure}

Electrostatic correlations are strong not only for artificial
systems involving DNA, but also for the DNA in the cell
\cite{CELL}.  In particular, to organize DNA in chromatin (in
eukariotic cells), nature uses proteins having large positive
charges - histones. Higher levels of chromatin hierarchical
structure can be disassembled, with some of the histones released,
by an elevated concentration of salt.  The resulting most stable
lower level structure is a bead-on-a-string necklace, as shown in
the Fig. \ref{fig:chromatin1}.  It is called 10 nm fiber, because
the beads, called nucleosomes, are about that big.  Each
nucleosome consists of a core particle, called an octamer, with a
total charge of about $+170 e$.  As shown in the inset in Fig.
\ref{fig:chromatin1}, the octamer is encircled 1.8 times by a DNA
having a charge $-292e$.  The electrostatic interaction energy
between the DNA molecule and the histone octamer dwarfs the
bending energy of the DNA molecule. Strikingly, a simple theory
discussed in Sec. \ref{sec:flexible} yields a similar structure,
Fig. \ref{fig:necklace}, based on purely electrostatic
correlations in a simple model.

Correlated screening has also many useful practical applications.
The first to be mentioned is the technology of dressing of DNA
with polycations (Kabanov and Kabanov, 1995 and 1998), positively
charged star-like polymers called dendrimers (Tang et al, 1996;
Kabanov et al, 2000; Evans et al, 2001), or liposomes (Radler et
al, 1997) in order to produce positive complexes with DNA. This
facilitates gene delivery through a negative cell surface membrane
\cite{GeneTherapy}. There is also the idea to manufacture
nano-wires by attaching positive silver or gold colloids to DNA
(Braun et al, 1998; Keren et al, 2001).

We view these examples as both important and convincing enough to
engage in the study of electrostatic correlations between strongly
charged $Z$-ions. Strong correlations manifest themselves in a
number of ways and alter dramatically the whole picture of
screening.  In the familiar Debye theory, screening somewhat
reduces the effective value of charge as seen from a finite
distance in the outside world. With strongly correlated ions,
over-screening becomes possible, in which case the  shielded
macroion charge is seen from outside as having the opposite sign.
This counterintuitive phenomenon is called {\em charge inversion}.

%

The first experimental study related to charge inversion was
reported a long time ago by De Jong (1949).  More recently, charge
inverted complexes of polyelectrolytes were directly observed in
electrophoresis experiments (see review articles by Kabanov and
Kabanov (1995 and 1998) and references therein).  It is now
understood that charge inversion is a generic phenomenon and it is
expected in all systems in which strongly charged ions participate
in screening. It turns out to be a natural manifestation of
correlations between screening ions. In recent years, the
phenomenon of charge inversion has attracted significant attention
of theorists (Ennis et al, 1996; Wallin and Linse, 1996; Perel and
Shklovskii, 1999; Shklovskii, 1999b; Mateescu et al, 1999; Park et
al, 1999; Joanny, 1999; Sens and Gurovich, 1999; Netz and Joanny,
1999a; Netz and Joanny, 1999b; Wang et al, 1999; Nguyen et al,
2000a-b; Messina et al, 2000; Nguyen and Shklovskii, 2001a-e;
Chodanowski and Stoll, 2001; Dobrynin et al, 2001; Andelman and
Joanny, 2000; Potemkin et al, 2001; Tanaka and Grosberg, 2001a.)

Another equally interesting manifestation of correlations is the
possibility of attraction between like charged macroions mediated
by $Z$-ions (Rouzina and Bloomfield, 1996; Gronbech-Jensen et al,
1997; Levin et al, 1999; Moreira and Netz, 2000b).  This has
implications in the large field of self-assembly of charged
biological objects, ranging from RNA (Woodson, 2000; Pan et al,
1999) to virus heads \cite{PhysicsToday}.  In Sec.
\ref{sec:forces}, we show how this attraction of like charges
competes with repulsion due to inverted charges, which combine to
induce reentrant condensation of DNA or colloids out of solutions.

Although the subject matter of our Colloquium belongs to chemical
and biological physics, it has remarkably many deep physical
analogies in other branches of physics, from atomic physics to
quantum Hall effect;  we discuss this in Sec. \ref{sec:analogies}.

In what follows, we re-examine the entire concept of screening to
include correlations.  Although this requires stepping beyond the
mean field approximation, we shall start with some historical
remarks which also serve to define the notations.

\section{Historical remarks:  Mean field
theories}\label{sec:history}

The history of our subject is almost one century long.  While we
are unable in this brief Colloquium to discuss it in any depth, we
shall mention three key steps.

In the first step, Gouy (1910) and independently Chapman (1913)
examined the double layer at an electrode surface, an intensely
disputed subject at the time.   Following Gouy and Chapman, let us
consider a massive insulating macroion.  To be specific, assume
that its charge is negative, with surface density $- \sigma$.
Assume further that the only counterions in the system are those
dissociated from the surface.  Since a macroion is large, the
problem of the counterion distribution near the surface is
one-dimensional, with both counterion concentration $N(x)$ and
electrostatic potential $\phi(x)$ depending on the distance $x$
from the surface.   Gouy and Chapman addressed this problem by
solving the following equation:
\be \Delta \phi = - \frac{4\pi}{\epsilon} N_s e Z \exp \left[ - e
Z \frac{ \phi (x) - \phi_s}{ k_B T} \right]  \ . \label{eq:PB}
\ee
Here $\epsilon \approx 80$ is the dielectric constant of water,
$\phi_s$ and $N_s$ are the potential and counterion concentration
at the surface.  Formula (\ref{eq:PB}) follows from the Poisson
equation for the potential $\phi (x)$ and the assumption that the
ions determining the charge density are Boltzmann-distributed in
the same potential.  Because of this self-consistency assumption,
this equation describes a mean field approximation.  It is
universally called the Poisson-Boltzmann equation.  One boundary
condition $\left. d \phi / dx \right|_{x=0} = 4 \pi \sigma /
\epsilon$ follows from the fact that the field vanishes on the
other side of the boundary (as there is no electrolyte and there
are no charges).  The second condition is that the concentration
$N(x)$ must be normalized to $\sigma / Z e$ ions per unit area.
The exact solution of the thus formulated Gouy-Chapman problem
reads
\be N(x) = \frac{k_B T \epsilon /2 \pi e^2 Z^2}{\left( x + \lambda
\right)^2} \label{GCsolution} \ ,
\ee
where $\lambda$ is called Gouy-Chapman length, which is equal to
\be \lambda = k_B T \epsilon / 2 \pi \sigma Z e \ .
\label{eq:lambda}
\ee
The interpretation of Eq. (\ref{eq:lambda}) is interesting. If a
$Z$-ion were to be alone next to the charged plane, it would be
confined by the surface field $2 \pi \sigma / \epsilon$ to such a
``height'' $\lambda$ that its energy change $2 \pi \sigma Ze
\lambda / \epsilon$ is about $k_B T$ - which leads to the correct
answer Eq. (\ref{eq:lambda}).  Other $Z$-ions cancel the field
inside the macroion but double the field in the electrolyte at the
macroion surface (see above the boundary condition at $x=0$).  On
the other hand, every particular ion at every moment is higher
than, roughly, half the other ions, whereupon it finds itself in a
partially screened field. Exact solution of the Poisson-Boltzmann
equation indicates that these two factors cancel each other,
yielding Eq. (\ref{eq:lambda}).

The next step of the screening story, well known to every
physicist, is the theory of Debye and H\"uckel (1923) initially
developed for electrolytes - the overall neutral mixture of mobile
ions of both signs, and now widely used in plasma and solid state
physics. Debye and H\"uckel linearized the Poisson-Boltzmann
equation (\ref{eq:PB}) (generalized by introducing the sum over
ion species on the right-hand side).  Of course, linearization can
be done if the potential is not too strong anywhere in the system,
which is often the case if charges involved are small enough.
Debye-H\"uckel screening leads to exponential decay of the
potential around a point-like charge $Q$:
\be \phi (r) = \frac{Q}{\epsilon r}e^{- r/r\!_s} \ , \ee
where the Debye-H\"uckel radius is given by
\be r\!_s = \left( \frac{k_B T }{ 8 \pi N_1 e^2} \right)^{1/2} \ ,
\label{eq:Debye_radius} \ee
and $N_1$ is the concentration of monovalent salt.

Less well known among physicists is the fact that Debye-H\"uckel
theory ignited a heated debate: Bjerrum (1926) commented that
$\int \exp \left[ e^2 / r k_B T \right] r^2 dr$ diverges at $r \to
0$ and, therefore, point-like charged particles are expected to
associate in neutral pairs.  The discussion led to the realization
of the important role of short range repelling forces.  In order
to prevent association of monovalent ions and formation of Bjerrum
pairs, the repulsion forces should take over at the distance which
is not much smaller than the so-called Bjerrum length
\be \ell_B = e^2/ \epsilon k_B T \ ,
\label{eq:Bjerrum} \ee
which is about 0.7 nm in water at room temperature.  (This is why
hydration layer of a few water molecules around each ion is
essential to stabilize dissociated ions).

The last step we mention here is relatively recent, it has to do
with non-linear screening of cylindrical charges, such as the DNA
double helix (Onsager, 1967; Manning, 1969; Oosawa, 1971).
Consider a cylinder charged to the linear density $-\eta$. Since
the potential is logarithmic, its competition with entropy is
quite peculiar. Indeed, releasing counterions to some distance $r$
requires energy $(2 e Z \eta / \epsilon) \ln (r/a)$ ($a$ being the
cylinder radius), while the corresponding entropy gain is $ k_B T
\ln (\pi r^2 / \pi a^2)$. Therefore, counterions are released only
as long as $\eta < \eta_Z$, where
\be \eta_Z = k_B T \epsilon / e Z \ .
\label{eq:OM_critical_charge} \ee
When a cylinder is charged in excess of $- \eta_Z$, some of the
ions remain {\em Onsager-Manning condensed} on the cylinder, so
that its effective net charge is equal to $- \eta_Z$. To emphasize
the importance of this subject, let us mention that the DNA double
helix has a bare charge density of about $-4.2 \eta_1$, where
$\left. \eta_1 = \eta_Z \right|_{Z=1}$ (see Frank-Kamenetskii et
al (1987) for further DNA applications). Onsager-Manning
condensation was more accurately justified by Zimm and Le Bret
(1983).  These authors addressed non-linear Poisson-Boltzmann
equation in cylindrical geometry (refining earlier works - see the
collection of papers Katzir-Katchalsky (1971)).

Gouy-Chapman, Debye-H\"uckel, and
Onsager-Manning theories are all of mean field type, based on the
Poisson-Boltzmann equation.  This approach works well when
screening charges are small, $Z=1$.  In the case of strongly
charged macroions and $Z$-ions with $Z \gg 1$, however,
correlations are important and mean field theory fails.  Hence it
is necessary to step beyond this approximation. This is precisely
the subject matter of the present Colloquium.  Charge inversion in
this context should be viewed as the most obvious manifestation of
the failure of the mean field approximation.

\section{Strongly correlated liquid of multivalent ions}\label{sec:small_ions}

\begin{figure}
\epsfxsize=5cm \centerline{\epsfbox
{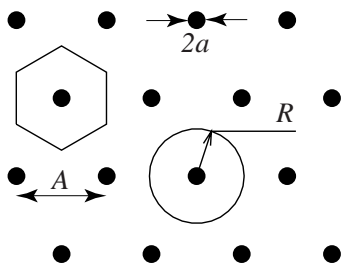}} \caption{A Wigner crystal of positive $Z$-ions
on a uniform background of negative surface charge. A hexagonal
Wigner-Seitz cell and its simplified version as a disk with radius
$R$  are shown.}\label{fig:WC}
\end{figure}

To begin with, let us explain why the Gouy-Chapman solution
(\ref{GCsolution}) fails at large $Z$. Apart from $\lambda$
(\ref{eq:lambda}), there is a second length scale in the problem
due to the discreteness of charge.  It is associated with the
distance between ions in the lateral direction, along the plane.
As long as the system as a whole is neutral, the two-dimensional
concentration of $Z$-ions is $n = \sigma / Ze$, and the surface
area per ion can be characterized by a radius $R$ such that $\pi
R^2 = 1/n$ (see Fig. \ref{fig:WC}).  Thus  $R = (\pi n )^{-1/2} =
( Z e / \pi \sigma )^{1/2}$, and hence (cf. Eq. (\ref{eq:lambda})
\be \frac{R}{\lambda} = 2 \Gamma \ , \ \ \ \Gamma = \frac{Z^2 e^2
/\epsilon R}{ k_B T} \ . \label{eq:Gamma} \ee
Here $\Gamma$ is the Coulomb coupling constant, or the inverse
dimensionless temperature measured in the units of a typical
interaction energy between $Z$-ions.  A system of monovalent ions,
$Z=1$, is weakly coupled, $\Gamma \sim 1$, this is why classical
mean field theory applies.  By contrast, a system in which
$Z$-ions have large $Z$ is strongly coupled, and we see that $R$
becomes larger than $\lambda$. For example, at $Z=3$ and $\sigma =
1.0~e/$nm$^{2}$ we get $\Gamma = 6.4$, $\lambda \simeq 0.1$ nm and
$R \simeq 1.0$ nm. Clearly, mean field treatment along the lines
of Poisson-Boltzmann theory fails in this situation, since
$Z$-ions do not affect each other when they are at distances
smaller than $R$ from the plane. It is worth emphasizing once
again that it is not only the {\em linearized} Debye-H\"uckel
theory that fails in the strong coupling regime, but so does also
the non-linear Poisson-Boltzmann theory.  It is the mean field
approximation that is inapplicable because of correlations between
discreet charges.

An alternative theory appropriate for the regime $\Gamma \gg 1$
was suggested by Perel and Shklovskii (1999).  The main idea of
this theory is that at $\Gamma \gg 1$ the screening atmosphere is
narrowly confined at the surface (see Fig. \ref{fig:WCpicture}),
and it should be approximated as a two-dimensional SCL.

A two-dimensional liquid of classical charged particles on a
neutralizing background, the so-called one component plasma, is
well understood \cite{Totsuji}.  At zero temperature, it acquires
the minimal energy state of a Wigner crystal, shown in the Fig.
\ref{fig:WC}, in which the correlation energy per ion and the
chemical potential are given by
\begin{eqnarray}
\varepsilon(n) & \simeq & - 1.11 Z ^{2}e^{2}/R  \epsilon   = -
1.96 n^{1/2}Z^{2}e^{2}/  \epsilon \ , \label{benergy} \\
\mu_{WC} & = & \frac{\partial[n \varepsilon(n)]}{\partial n}  =
\frac{3}{2} \varepsilon (n) = -1.65 {Z^{2}e^{2}}/{  \epsilon  R} \
. \label{eq:muwc}
\end{eqnarray}
We interpret $R$ here as the radius of a Wigner-Seitz cell
approximated by a disc (see Fig. \ref{fig:WC}).  

At non-zero temperature, the chemical potential of one component
plasma can be written as $\mu = \mu_{id} + \mu_{WC} + \delta \mu$.
Here $\mu_{id}$ is the chemical potential of an ideal gas at the
same concentration. Accordingly, $\mu_{WC} + \delta \mu$ part is
entirely due to correlations. Furthermore, it turns out that
$\delta \mu$, which is the thermal correction, is negligible for
$\Gamma \gg 1$. Although a Wigner {\em crystal}, in terms of long
range order, melts at $\Gamma \approx 130$, the value of $\delta
\mu$ is controlled by short range order and remains negligible as
long as $\Gamma \gg 1$.  It is in this sense that a SCL of
$Z$-ions is similar to a Wigner crystal.

\begin{figure}
\epsfxsize=6cm \centerline{\epsfbox
{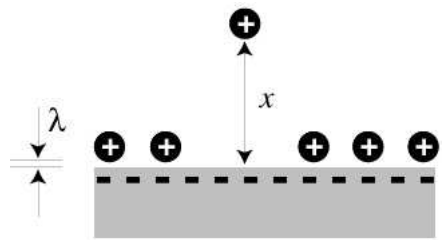}} \vspace{-5cm} \caption{The origin of attraction of
a new positive $Z$-ion to the already neutralized surface.
$Z$-ions are shown by solid circles. The new $Z$-ion creates its
negative correlation hole.}\label{fig:coming_ion}
\end{figure}

Thus, the correlation part of the chemical potential can be
approximated by $\mu_{WC}$ (\ref{eq:muwc}), which is negative and
large: $- \mu_{WC} /k_B T = 1.65 \Gamma \gg 1$.  The physics of a
large and negative $\mu_{WC}$ can be understood as follows.
Pretend for a moment that the insulating macroion is replaced by a
neutral metallic particle. In this case, each $Z$-ion creates an
image charge of opposite sign inside the metal.  The energy of
attraction to the image is $U(x) = -(Ze)^2 / 4 \epsilon x$, where
$x$ is the distance to the surface.  This energy is minimal when
the $Z$-ion is placed next to the surface, at a distance equal to
its radius $a$; therefore, the $Z$-ion sticks to the surface. With
this idea in mind, consider bringing a new $Z$-ion to the
insulating macroion surface already covered by an adsorbed layer
of $Z$-ions (Fig. \ref{fig:coming_ion}). This layer behaves like a
metal surface in the sense that the new $Z$-ion repels adsorbed
ones, creating a correlation hole.  In other words, it creates a
negative image. Because of the discreteness of charges, the
adsorbed layer is a good metal at length scales above $R$ only.
Accordingly, attraction to the image gets saturated at $x \sim R$.
This is why the chemical potential of an ion in a Wigner crystal
scales as $\mu_{WC} \sim -(Ze)^{2}/ \epsilon R$. Eq.
(\ref{eq:muwc}) specifies the numerical coefficient in this
expression.

Using the concept of images, we can now understand the
distribution of $Z$-ions, $N(x)$, near the surface.  To do this,
let us extract one $Z$-ion from the SCL and move it along the $x$
axis. As long as $ x \ll R$, its correlation hole does not change,
and, therefore, the $Z$-ion is attracted to the surface by the
uniform electric field $E = 2 \pi \sigma/ \epsilon $; other
$Z$-ions do not affect this attraction in any way.  Therefore,
$N(x)= N_s \exp(-x/\lambda)$ for $x \ll R$. Here $N_s \simeq
n/\lambda$ is the three-dimensional concentration of $Z$-ions
close to the surface plane (Eq. (\ref{eq:PB})). For $x \gg R$, the
correlation hole acts as a point-like image charge, the
corresponding interaction energy being $-Z^2 e^2 / 4 \epsilon x $.
At $x=Z^2e^2/4 \epsilon k_B T = R_0 \Gamma / 4 = \lambda \Gamma^2
/ 2$, the interaction with the image charge drops to about $k_B
T$, i.e., negligible, and the $Z$-ion concentration becomes
\begin{equation}
N_0 = N_s \exp \left(- \frac{|\mu_{WC}|}{k_B T} \right) = N_s \exp
\left( - \frac{ 1.65 Z^2 e^2}{ \epsilon R k_B T} \right) \ .
\label{eq:N0}
\end{equation}
We shall further comment on the physical meaning of $N_0$ after
Eq. (\ref{eq:Qstar}).  Note that the correction term $- Z^2 e^2 /
4 \epsilon x$ to $Z$-ion energy, which is important in the
interval $R \ll x \ll \lambda \Gamma^2 /2$, is similar to the
``image'' correction to the work function of a metal \cite{Lang}.

The dramatic difference between the exponential decay of $N(x)$
and the Gouy-Chapman $1/(\lambda + x)^2$-law (\ref{GCsolution}) is
due to correlation effects. Moreira and Netz (2000a) re-derived
these results in a more formal way and confirmed them by
Monte-Carlo simulations.  Recently, Moreira and Netz (2001) also
showed that discreteness of the surface charge neglected above
leads to the lateral pinning of $Z$-ions.  This brings $Z$-ions
even somewhat closer to the surface. General direction of this
effect can be understood from the limit (although unrealistic for
strongly charged surface) when the distance of closest approach
between a discrete surface charge and a $Z$-ion is so small that
they  form isolated Bjerrum pairs (see (\ref{eq:Bjerrum})).

At larger $x$, correlations and interactions with image charges
are unimportant, and the Poisson-Boltzmann equation applies.  In
this region $N(x)$ varies so smoothly that $N(x) = N_0$ provides
an effective boundary condition for the Poisson-Boltzmann equation
\cite{Shklov99}.  At this stage we must remember that in a real
physical situation there is always some concentration of $Z$-ions,
$N$, in the surrounding solution.  It can be either larger or
smaller than $N_0$.  In the latter case, the surface is
overcharged, as we show in Sec. \ref{sec:inversion} below.

\section{Correlation-induced charge inversion}\label{sec:inversion}

Let us return once again to the physical argument illustrated by
Fig. \ref{fig:coming_ion}.  It explains why an extra $Z$-ion may
be attracted to the macroion surface despite the fact that it is
already neutralized by the previously adsorbed $Z$-ions.  What
happens if another $Z$-ion approaches?  Clearly, correlation
effect will keep providing an attractive force for this and
subsequent $Z$-ions, but it will have to compete with the
repulsive force which is simply due to the fact that macroion has
already too many $Z$-ions adsorbed and the whole complex is,
therefore, positively charged.  Thus, the question is this:  what
is the {\it equilibrium} amount of (over)charge?

One useful way to think about it is to realize that correlation
mechanism provides voltage to drive overcharging, but the actual
amount of (over)charge depends on both voltage and capacitance.
Since the latter depends strongly on the geometry, we will have to
explore several cases - spherical macroion, cylindrical, etc.

Another, equivalent, view, involves the comparison of chemical
potentials of adsorbed $Z$-ions and $Z$-ions in the bulk solution.
This approach immediately suggests that in equilibrium the total
charge, $Q^{*}$, depends on the concentration, $N$, of $Z$-ions in
the surrounding bulk solution.  Here $Q^{*}$ is the net charge of
the entire complex which includes bare charge of the macroion,
$-Q<0$, and the proper (determined by the equilibrium condition)
number of adsorbed $Z$-ions with charge $Ze>0$ each.

Let us see now how we can implement the condition of equal
chemical potentials for the case of a spherical macroion, with
radius $r$. As regards adsorbed $Z$-ions, we argue\footnote{The
argument goes as follows.  Spherical surface of the macroion has
bare surface charge density $-\sigma = -Q / 4 \pi r^2$. Let us
pretend to place there, along with real charge $-\sigma$, also two
imaginary spheres, with uniform charge densities $\sigma^{*}
=Q^{*}/4 \pi r^2$ and $-\sigma^{*}$. Their total charge is zero,
so they have no effect.  However, we can now think that the
$Z$-ions are adsorbed on the sphere with the charge density
$-\sigma - \sigma^{*}$, and with this sphere they form a {\em
neutral} SCL, quite like that considered in Sec.
\ref{sec:small_ions}.  The remaining sphere has the charge density
$\sigma^{*}$ and creates spherically symmetric field with
potential on the surface $\psi (0)$.  We emphasize that the
macroscopic net charge $\sigma^{*}$ does not interact with the one
component plasma, because the potential $\psi(0)$ is constant
along the surface, while one component plasma is neutral.} that
their chemical potential is only different from the one-component
plasma expression (\ref{eq:muwc}) by the energy of the $Z$-ion in
the potential $\psi(0)=Q^{*}/\epsilon r$ created by the net charge
$Q^{*}$ . Therefore, equilibrium condition reads $\mu_{id} +
\mu_{WC} + Ze \psi (0) = \mu_b$, where $\mu_b$ is the bulk
chemical potential. To determine $Q^{*}$ from here, we first note
that $\mu_{id} - \mu_{b} = k_BT \ln(N_s / N)$; we further express
$\mu_{WC}$ in terms of $N_0$ (see Eq. (\ref{eq:N0})), and finally
obtain
\be Q^{*} = \frac{\epsilon r }{Ze} k_B T \ln (N / N_0) \ .
\label{eq:Qstar} \ee
Clearly, the net charge $Q^{*}$ is indeed positive when $N
> N_0$, i.e. it has the sign opposite to the bare charge
$Q$.

The result (\ref{eq:Qstar}) sheds also light on the meaning of the
quantity $N_0$ defined above, in Eq. (\ref{eq:N0}):  this is the
concentration of $Z$-ions in the surrounding bulk solution at
which macroion is exactly neutralized by the adsorbed $Z$-ions.
This concentration is very small because $|\mu_{WC}|/k_BT \gg 1$.
For example, $N_0 =0.3$~mM and $0.8 \ \mu$M for $Z = 3$ and 4
respectively ($1$~M~$ \approx 6 \times 10^{20}$cm$^{-3}$).
Therefore, it is easy to increase charge inversion by increasing
$N$.  How far does it go? At large enough $N$, translational
entropy terms $\mu_{b}- \mu_{id}$ become negligible compared to
$\mu_{WC}$, yielding
\begin{equation}
Q^{*}/\epsilon r = \psi(0) = |\mu_{WC}|/Ze \ . \label{capacitor}
\end{equation}
Expressing $R$ and $|\mu_{WC}|$ through $Q$ and $Z$ with the help
of Eq. (\ref{eq:muwc}) (and remembering that $\sigma^{*} = -
\sigma + Z en$), Shklovskii (1999b) arrived at the prediction for
the {\em maximal} inverted charge for the spherical macroion which
can be achieved by increasing concentration of $Z:1$ salt:
\begin{equation}
Q^{*} = 0.83\sqrt{Q Z e}. \label{eq:Q}
\end{equation}
This charge is much larger than $Ze$, but still is smaller than
$Q$ because of limitations imposed by the large charging energy.
For example, for $Q=100e$, $Z=4$, we get $Q^{*} = 17e$. Eq.
(\ref{eq:Q}) was recently confirmed by numerical simulations
(Messina et al, 2000; Tanaka and Grosberg, 2001a).  Further
increase of charge inversion beyond the level dictated by Eq.
(\ref{eq:Q}) is achievable with the help of a monovalent salt (see
Sec \ref{sec:salt}).

Applied in cylindrical geometry, similar arguments lead to a
revision of the conventional Onsager-Manning condensation theory
(Sec. \ref{sec:history}) when dealing with multivalent $Z$-ions.
Consider a cylinder with a negative linear charge density $ -
\eta$ and assume that $\eta > \eta_{Z}$ . Mean field
Onsager-Manning theory (\ref{eq:OM_critical_charge}) predicts
$\eta^{*}= -\eta_{Z}$. By contrast, Perel and Shklovskii (1999)
showed that a correlation-induced negative chemical potential
$\mu_{WC}$ results in
\begin{equation}
\eta^{*} = - \eta_{Z} \frac {\ln(N_0/N)}{\ln \left(  4  / \pi
Z^{6} N \ell_B^3 \right)} \ , \label{eq:etaapp}
\end{equation}
where $\ell_B$ is the Bjerrum length (\ref{eq:Bjerrum}).  This
result reproduces the Onsager-Manning one
(\ref{eq:OM_critical_charge}) only at extremely small values of
$N$, which are unrealistic at $Z \geq 3$. On the other hand, at $N
= N_0$ the net charge flips sign, resulting in charge inversion at
$N> N_0$ (which is absent in Onsager-Manning theory). At large
enough $N$, the inverted charge density $\eta^{*}$ can reach $k_B
T \epsilon / e = \eta_1$.

\section{Enhancement of charge inversion by monovalent
salt}\label{sec:salt}

Most water solutions, particularly biological ones, contain
significant amounts of monovalent salt, such as NaCl. Correlations
between these monovalent ions are negligible, and, therefore,
their only role is to provide Debye-H\"uckel screening with a
decay length $r\!_s$ (\ref{eq:Debye_radius}).  This screening
makes charge inversion substantially stronger.  Indeed, screening
by a monovalent salt diminishes the charging energy of the
macroion much stronger than the correlation energy of $Z$-ions.
Furthermore, in a sufficient concentration of salt, the macroion
is screened at the distance smaller than its size. Then, the
macroion can be thought of as an over-screened surface, with
inverted charge $Q^{*}$ proportional to the surface area.  In this
sense, overall shape of the macroion is irrelevant, at least to a
first approximation. Therefore, we consider here a simpler case:
screening of a planar macroion surface with a negative surface
charge density $-\sigma$ by a solution with a concentration $N$ of
$Z:1$ salt and a large concentration $N_1$ of a monovalent salt.

Nguyen et al (2000a) calculated analytically the dependence of the
charge inversion ratio, $\sigma^{*}/\sigma$, on $r\!_s$, in two
limiting cases $r\!_s\gg R_0$ and $r\!_s\ll R_0$, where $R_0 =
(\pi \sigma/ Ze)^{-1/2}$ is the radius of a Wigner-Seitz cell at
the neutral point $n=\sigma/Ze$.  At $r\!_s\gg R_0$ calculation
starts from Eq. (\ref{capacitor}).  The electrostatic potential of
a plane with the charge density $\sigma^{*}$ screened at the
distance $r\!_s$ reads $\psi(0) = 4 \pi \sigma^{*}r\!_s$. At
$r\!_s \gg R_0$ screening by monovalent ions does not change Eq.
(\ref{eq:muwc}) substantially so that we still can use it in Eq.
(\ref{capacitor}) which now describes charging of a plane
capacitor by voltage $|\mu_{WC}|/Ze$. This gives
\begin{equation}
\sigma^{*}/\sigma =  0.41 (R_0/r\!_s) \ll 1 \ \ \ \ \ (r\!_s \gg
R_0) \ . \label{smallyI}
\end{equation}
Thus, at $r\!_s \gg R_0$ inverted charge density grows with
decreasing $r\!_s$.

Now we switch to the case of strong screening by monovalent salt.
To begin with, let us assume that screening is already so strong
that $r\!_s \ll R_0$, but energy of SCL is still much greater than
$k_BT$ per $Z$-ion.  In this regime, free energy consists of
Debye-H\"uckel screened nearest neighbor repulsion energies of
$Z$-ions and the attraction energy of $Z$-ions to the charged
surface:
\begin{equation}
F=   (3nZ^{2}e^{2}/  \epsilon A)\exp(-A/r\!_s) - 4 \pi \sigma
r\!_s Zen/  \epsilon \ , \label{SCWC}
\end{equation}
where $A = (2/\sqrt3)^{1/2}n^{-1/2}$ is the lattice constant of
the hexagonal Wigner crystal  (Fig. \ref{fig:WC}). Minimizing $F$
with respect to $n$ one arrives at
\begin{equation}
\frac{\sigma^{*}}{\sigma } = \frac{\pi }{ 2\sqrt{3}} \left( \frac{
R_{0} }{ r\!_{s} \ln(R_{0}/r\!_{s}) }\right)^{2} \ \ \ \ \ \
(r\!_s \ll R_0) \ . \label{giantI}
\end{equation}
Thus $\sigma^{*}/\sigma$ grows with decreasing $r\!_s$ and can
become larger than 100\%. At $r\!_s \sim R_0$, Eq. (\ref{smallyI})
and Eq. (\ref{giantI})  match each other.  As we see, $\sigma^{*}$
continues to grow with decreasing $r\!_s$.  This is because the
repulsion between $Z$-ions becomes weaker, so that it is easier to
pack more $Z$-ions on the surface.  Of course, when $r\!_s$
decreases even further, binding energy of $Z$-ions becomes small,
SCL dissolves, and charge inversion disappears.

\begin{figure}
\epsfxsize=6cm \centerline{\epsfbox
{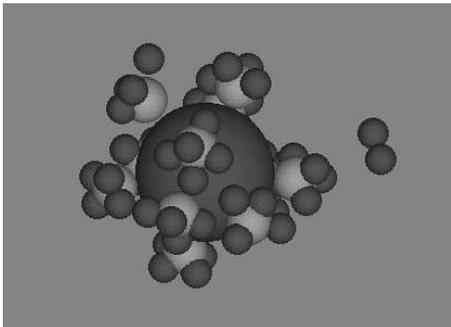}} 
\caption{The snapshot of the system simulated by Tanaka and
Grosberg (2001a). The spherical macroion is grey, $Z$-ions are
white, and monovalent negative ions are black.
 Bare charge of the spherical macroion is
$Q=-28e$, $Z=7$, and $\Gamma \approx 30 $. There are 52 $Z$-ions
and 336 monovalent ions in the simulation domain.  What is clearly
seen is the formation of $Z^{\prime}$-ions:  because $Z$ is so
large, $Z$-ions adsorb monovalent ions.  This reduces correlations
between $Z$-ions and restricts charge inversion.  Nevertheless,
bare charge of the complex shown in the figure is $+16 e$.
}\label{fig:Motohiko}
\end{figure}

The results above are in a good agreement with simulations.  Terao
and Nakayama (2001) reported the results of a Monte Carlo
simulation for the system consisting of a macroion with charge
$Q=20e$ surrounded by 20 monovalent counterions and 1500 ions of
$2:1$ electrolyte. Tanaka and Grosberg (2001a) performed molecular
dynamics simulations with a spherical macroion of the charge
$Q=-28e$, spherical $Z$-ions ($2 \leq Z \leq 7$), and up to 500
monovalent ions; the system is neutral overall, which determines
the number of $Z$-ions to be between 180 and 52.  Simulations
confirmed the strong adsorption of the overcharging amount of
$Z$-ions on the surface of macroion. On a more quantitative level,
Tanaka and Grosberg (2001a) examined the dependence of the
inverted charge on the ionic strength, $I$, and found the
crossover between $Q^{*} \propto \sqrt{ I} \propto 1 / r\!_s$ and
$Q^{*} \propto I \propto 1 / r\!_s^2$, consistent with Eqs.
(\ref{smallyI}) and (\ref{giantI}). Tanaka and Grosberg (2001a)
attempted also to maximize the charge inversion ratio $Q^{*} / Q$.
In agreement with the theoretical views presented above, the
growth of charge inversion is capped when correlations between
$Z$-ions are suppressed. We have mentioned eventual reduction of
correlations for large monovalent salt concentration, or small
$r\!_s$, when the SCL evaporates.  If one tries to increase $Z$
instead of lowering $r\!_s$, then, at large $Z$, correlations
become suppressed because monovalent ions condense on $Z$-ions,
forming $Z^{\prime}$-ions with smaller net charge $Z^{\prime}$.
This effect is clearly seen in Fig.\ref{fig:Motohiko} (see Nguyen
et al (2000b) for the conditions under which this phenomenon is
and is not important). Nevertheless, $Q^{*}/ Q$ up to about 150\%
is easily observed.

\section{Screening of a charged plane by polyelectrolytes}\label{sec:rods}

\begin{figure}
\epsfxsize=6.5cm \centerline{\epsfbox{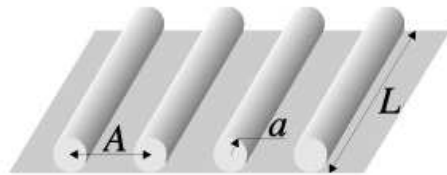}} \caption{Rod-like
negative $Z$-ions, adsorbed on a positive uniformly charged
plane.}\label{fig:rods}
\end{figure}

A practically important class of $Z$-ions are charged polymers,
i.e., polyelectrolytes.  Let us start with a rigid polyelectrolyte
and discuss charge inversion caused by adsorption of long rod-like
$Z$-ions.  For example, the moderately long (up to about $50$ nm,
or about 150 base pairs) DNA double helix can be well approximated
as a rod. Actin is another example of an even more rigid
polyelectrolyte.  Apart from the uninteresting regime of extremely
small macroion surface charge density (in which case the elongated
shape of molecules is irrelevant, rendering our previous results
applicable), charged rods adsorbed at the surface tend to be
parallel to each other due to the strong lateral repulsion.  In
other words, there is the short range order of a one-dimensional
Wigner crystal with lattice constant $A$ in the direction
perpendicular to the rods (Fig. \ref{fig:rods}).

To make the signs consistent with the case of DNA, we assume that
the polyelectrolyte charge is negative and equal to $- \eta$ per
unit length, while the macroion surface is a plane with positive
charge density $\sigma$.  We assume also that there is a certain
concentration of monovalent salt, $N_1$, in the solution,
corresponding to the Debye screening radius
(\ref{eq:Debye_radius}).  To begin with, let us assume that charge
density of the rods, $-\eta$, is below the Onsager-Manning
threshold, Eq. (\ref{eq:OM_critical_charge}), and let us apply the
Debye-H\"uckel approximation to describe the screening of the
charged surface by the monovalent salt.  We can then directly
minimize the free energy of the one-dimensional crystal of
negative rods on the positive surface written similarly to Eq.
(\ref{SCWC}). Then the competition between the attraction of the
rods to the surface and the repulsion of the neighboring rods
results in the negative net surface charge density $-\sigma^{*}$
similar to Eq. (\ref{giantI}) (Netz and Joanny, 1999a; Nguyen et
al, 2000a):
\begin{equation}
\frac{ \sigma^{*}}{\sigma } = \frac{ \eta / \sigma r\!_s }{
\ln(\eta / \sigma r\!_s ) } \ \ , \   r\!_s \ll A_0  \ .
\label{giant2}
\end{equation}
Here the applicability condition involves $A_0 = \eta / \sigma$,
which is the distance between rods when they neutralize the plane;
only at $r\!_s \ll A_0$ is the overcharged plane linearly screened
by monovalent salt.

Speaking about DNA, we have already discussed in Sec.
\ref{sec:history} that the DNA charge density, $-\eta$, is such
that about three quarters of it is compensated by positive
Onsager-Manning-condensed monovalent ions. In other words, the net
charge of DNA in the bulk solution is $\eta^{*} = -\eta_1$
(\ref{eq:OM_critical_charge}).  It turns out that at $r\!_s \ll
A_0$ the result (\ref{giant2}) applies to DNA with the only
correction of replacing $-\eta$ with $\eta^{*} = -\eta_1 \equiv
-k_B T \epsilon /e $ \cite{Shklov001}.

Thus the inversion ratio grows with decreasing $r\!_s$ as in the
case of spherical $Z$-ions.  At small enough $r\!_s$ and $\sigma$,
the inversion ratio can reach 200\% before DNA molecules are
released from the surface. It is larger than for spherical ions,
because in this case, due to the large length of the DNA helix,
the correlation energy remains large and the Wigner crystal-like
short range order is preserved at smaller values of $\sigma
r\!_s$. Nguyen et al (2000a) called this phenomenon ``giant charge
inversion.''

Let us switch now to the opposite extreme of weak screening by a
monovalent salt, $r\!_s \gg A_0$.  In this case, screening of the
overcharged plane by monovalent salt becomes strongly nonlinear,
with the Gouy-Chapman screening length $\lambda^{*} =  \epsilon
k_{B}T/ (2\pi e \sigma^{*})$ much smaller than $r\!_s$.
Furthermore, some of the positive monovalent ions Onsager-Manning
condensed on DNA are released from it upon adsorption, as the
plane repels them (Park et al, 1999; Gelbart et al, 2000). As a
result the absolute value of the net linear charge density of each
adsorbed DNA, $\eta^{*}$, becomes larger than $\eta_1$.  To
determine $\sigma^{*}$ and $\eta^{*}$, Nguyen et al (2000a)
considered two equilibrium conditions, dealing with chemical
potentials of rods and small ions, respectively.  As a result,
they arrived at the following beautiful formulae valid at $r\!_s
\gg A_0$:
\begin{equation}
\frac{\sigma^{*}}{\sigma} = \frac{\eta_1 }{ 2\pi a \sigma} \exp
\left( - \sqrt{  \ln \frac{ r\!_s }{ a } \ln  \frac{ A_0 }{ 2\pi a
}}  \right) \ , \label{NL1}
\end{equation}
\begin{equation}
\eta^{*}=  \eta_1  \sqrt{ \frac{\ln (r\!_s/ a)}{ \ln (A_0/2\pi
a)}} \ , \ \ \  r\!_s \gg A_0 \ . \label{NL2}
\end{equation}
At $r\!_s \simeq A_0 / 2 \pi $ we get $\eta^{*} \simeq \eta_1$,
$\lambda^{*} \simeq r\!_s$ and $\sigma^{*}/\sigma \simeq
\eta_1/(2\pi r\!_s \sigma)$ so that Eq.~(\ref{NL1}) crosses over
smoothly to the strong screening  result of Eq.~(\ref{giant2}).

So far in this Section we assumed a rod-like polyelectrolyte. Let
us now discuss how chain flexibility affects charge inversion. We
argue that the results for rod-like $Z$-ions are remarkably
robust.  To begin with, consider a polyelectrolyte having several
charged groups per each persistence length.  We argue that our
results remain valid as long as adsorption energy {\em per one
persistence length} is large compared to $k_B T$. Indeed, under
this condition even flexible polyelectrolyte chains lay flat on
the surface, in which case they are ordered in Wigner crystal-like
SCL and, therefore, behave similarly to rods.

Dobrynin et al (2001) addressed the opposite extreme, namely,
weakly charged polyelectrolytes, with so small a fraction of
charged monomers, $f$, that a link between two neighboring charges
is already a flexible polymer; in other words, the distance
between charges is larger than the persistence length.  It was
discovered by de Gennes et al (1976) that a weakly charged
polyelectrolyte chain in a bulk solution consists of electrostatic
blobs. Inside each blob polymer is only marginally perturbed by
Coulomb interactions, while chain of blobs is fully stretched,
rod-like. Dobrynin et al (2001) have found that this blob
structure remains valid for the adsorbed chains, which form SCL of
effective rods of blobs. Speaking of charge inversion, this means
that Eq. (\ref{giant2}) remains valid for the weakly charged
chains, provided $\eta$ is replaced with a linear charge density
of the string of blobs $\eta^{*} = (f \epsilon e k_B T /
l^2)^{1/3}$, where $l$ is the chain persistence length, and
$r\!_s$ is larger than a blob size.

With increasing $\sigma$, adsorbed rods, either real or blob ones,
start to touch each other leading to multilayer adsorption.  It is
only in this regime that the real and blob rods behave
differently, as we discuss in Sec. \ref{sec:layers}.

\section{Polyelectrolytes wrapping around charged particles}\label{sec:flexible}

\begin{figure}
\epsfxsize=7cm \centerline{\epsfbox {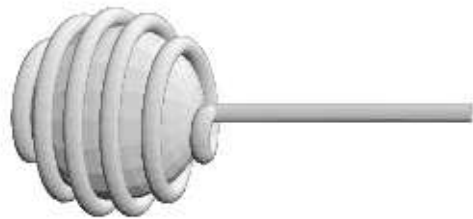}} \caption{A
polyelectrolyte molecule winding around a spherical macroion. Due
to the Coulomb repulsion, neighboring turns, which play the role
of $Z$-ions, are strongly correlated.}\label{fig:one_sphere}
\end{figure}

As we have mentioned in the Introduction, one of the practically
important situations is that of a long charged polymer forming
complexes with oppositely charged particles.  Motivated by the
problem of nucleosome charge inversion, Mateescu et al (1999),
Park et al (1999), Sens and Gurovich (1999), and Netz and Joanny
(1999b) considered the complex of a positive sphere with charge
$q$ and a negative polyelectrolyte, such as a DNA double helix,
which has to make some $n_t > 1$ turns around the sphere to
neutralize it (Fig. \ref{fig:one_sphere}). These authors predicted
a substantial charge inversion: more of the polyelectrolyte is
wound than is necessary to neutralize a sphere. Furthermore,
Mateescu et al (1999) found that a tightly coiled polyelectrolyte
conformation becomes unstable when the chain length exceeds a
certain threshold, and then an almost straight tail abruptly
stretches out (Fig. \ref{fig:one_sphere}).

Nguyen and Shklovskii (2001a) emphasized the role of correlations
in this case of charge inversion.  Indeed, neighboring turns repel
each other and form an almost equidistant solenoid, which locally
resembles SCL. The tail of the polyelectrolyte repels the already
adsorbed part of the polyelectrolyte and creates a correlation
hole, which attracts the tail back to the surface (compare Fig.
\ref{fig:coming_ion}). As a result, the net charge of the sphere
with wrapped polyelectrolyte, $q^{*}$, is negative. It is shown
that at $r\!_s \to \infty$ the charge inversion ratio scales as
$|q^{*}|/q \sim 1/n_t$.  On the other hand, at small enough
$r\!_s$ it can exceed 100\%.

Even more interesting is the system in which charged polymer is so
long that it forms complexes with many oppositely charged
particles (see Fig. \ref{fig:necklace}).  Examples include
micelles \cite{Dubin}, globular proteins (Kabanov et al, 1976; Xia
and Dubin, 1994), colloids (Braun et al, 1998; Keren et al, 2001),
or dendrimers (Kabanov et al, 2000; Evans et al, 2001), and, last
but not least, histone octamers forming 10 nm chromatin fiber with
DNA (Fig. \ref{fig:chromatin1}).  To be specific, we remain with
the signs consistent with the DNA case and consider a long
negative polymer chain in the solution of positive spheres. If the
concentration of spheres is large, it is favorable to adsorb many
spheres on the polyelectrolyte chain. As a result, each sphere is
under-screened by polyelectrolyte and has positive net charge.
Then, adsorbed spheres repel each other and the complex forms a
periodic necklace (see Fig. \ref{fig:necklace}). This necklace is,
in fact, a one-dimensional Wigner crystal, or SCL, of spheres,
which serve as $Z$-ions . Indeed, since the segment of the
polyelectrolyte wound around one sphere interacts almost
exclusively with this sphere, it plays the role of the
Wigner-Seitz cell.  Because of correlations, spheres bind to
polyelectrolyte in such a large number that the net charge of the
polyelectrolyte molecule becomes positive \cite{Shklov003}. In
this case, the charge inversion ratio scales as $Q^{*}/Q \sim
n_t^{1/4}$ in the absence of monovalent salt, where $-Q$ and
$Q^{*}$ are the bare and net charges of the polyelectrolyte
molecule, respectively. This means that charge inversion may be
larger than 100\%.  As we discussed in Sec. \ref{sec:salt}, charge
inversion can be further enhanced by a monovalent salt, in which
case $Q^{*}/Q \sim n_t$.  We shall return to the complexes of a
charged chain with spheres in Sec. \ref{sec:forces}.

\section{Multilayer adsorption}\label{sec:layers}

\begin{figure}
\epsfxsize=7.5cm \centerline{\epsfbox{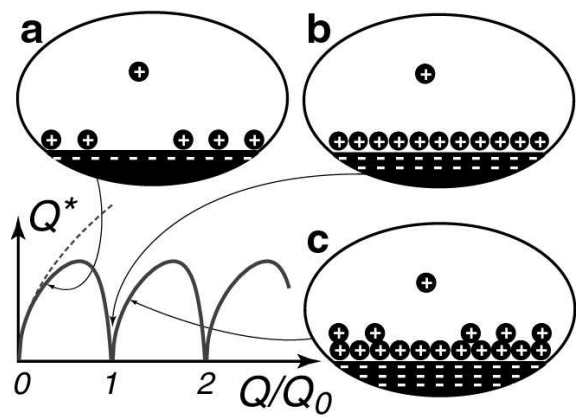}} \vspace{-3cm}
\caption{Inverted charge $Q^{*}$ as a function of the absolute
value $Q$ of the bare charge; $Q_0$ is the charge of one full
layer of $Z$-ions.  The dashed line corresponds to the case of
$Z$-ions with vanishing radius, Eq. (\protect\ref{eq:Q}).  (a) The
first layer is not full, as in Fig. \protect\ref{fig:coming_ion}.
An approaching new ion creates a correlation hole and is attracted
to it. (b) The layer is full, there is no place for a correlation
hole. (c) More than one layer is full. A correlation hole exists
in the top layer only. } \label{fig:layer}
\end{figure}

So far we have not considered the possibility that $Z$-ions fully
cover the macroion surface.  This may sometimes happen,
particularly when a macroion is very strongly charged, or $Z$-ions
are large. Suppose, for instance, that each $Z$-ion has a hard
core of radius $a$.  In this case, the effect of excluded volume
of the hard cores of the $Z$-ions adds positive contributions to
the surface pressure and chemical potential of a SCL
(\ref{eq:muwc}) that are proportional to $k_BT$ and that diverge
at the full coverage. Close to the full layer this term
compensates and then over-compensates the negative Coulomb term
$\mu_{WC}$, so that charge inversion disappears.  Indeed, a full
layer is incompressible (see Fig. \ref{fig:layer}b), and, unlike a
partially filled layer (see Fig. \ref{fig:coming_ion} or Fig.
\ref{fig:layer}a), it does not allow for the creation of an
image-like correlation hole.

At even larger macroion charge, the second layer starts to form,
launching a new wave of charge inversion. In the beginning, charge
inversion is small because all the attraction of a new $Z$-ion
approaching the surface is provided by a weak interaction with an
inflated image in the emerging second layer, where once again $A
\gg a$ (Fig. \ref{fig:layer}c). Continuing, Nguyen and Shklovskii
(2001c) arrived at the prediction of an oscillating inverted
charge $Q^{*}$ as a function of $Q$ (see Fig. \ref{fig:layer}),
where charge inversion vanishes every time the top layer of
$Z$-ions is full.

Another way to look at this phenomenon is to examine a metallic
electrode screened by $Z$-ions, when the potential of the
electrode is controlled instead of its charge. In this case,
oscillations of charge inversion and compressibility lead to
oscillations of capacitance of this electrode with the number of
adsorbed layers of $Z$-ions. This is similar to oscillations of
compressibility and magneto-capacitance in the quantum Hall
effect, which are related to consecutive filling of Landau levels
(Efros, 1988; Kravchenko et al, 1990; Eisenstein, 1992). In this
sense, we deal with a classical analog of the quantum Hall effect.

To conclude this section, let us return to the adsorption of
weakly charged polyelectrolytes that we discussed briefly in the
Sec. \ref{sec:flexible}. Dobrynin et al (2001) have shown that
parallel chains of adsorbed blobs start touching each other above
the same threshold surface charge density $\sigma_e = e f/ l^2$,
which corresponds to the onset of squashing blobs on the surface.
As a result, these authors arrived at the conclusion that if
weakly-charged chains are adsorbed on the surface with $\sigma
> \sigma_e$, they form a polymer liquid. In this liquid, correlations and image
formation are only due to the uppermost layer, with the thickness
about that of an unperturbed blob. There are no oscillations of
inverted charge; instead, charge inversion saturates at about one
layer of blobs and remains unchanged afterwards.

\section{Correlation-induced attraction of like
charges}\label{sec:forces}

The idea of a single screened macroion is a useful one in a
theoretical context, but in practice it is rarely true that there
is only one macroion.  Typically, there is a certain concentration
of them, so that interactions between them can be important.  Let
us start with the simplest question:  consider two macroions, and
suppose the concentration of $Z$-ions in solution is equal to
$N_0$ (see Eq. (\ref{eq:N0})), such that each macroion forms a
neutral complex with $Z$-ions. How do these two neutral complexes
interact?  It turns out that they attract each other at short
distances and, therefore, tend to coagulate.  In other words, two
macroions of the same charge may attract each other because of the
presence of $Z$-ions. In general, this attraction of like charges
is as interesting a manifestation of correlations as charge
inversion, even though our present Colloquium emphasizes the
phenomenon of charge inversion. Nevertheless, we must discuss
attraction at least briefly, in order to prepare the ground for
the subsequent discussion of experiments (Sec.
\ref{sec:experiment}).

Medium-induced attraction of like charges is nothing new in
physics, with Cooper pairs of electrons being the most prominent
example.  In the context of molecular physics, the most popular
explanation of attraction is in terms of {\em salt bridges}: a
divalent ion, such as Mg$^{2+}$, can form ionic bonds with two
groups with charges of $-1$ each, in effect connecting them
together. This idea is indeed adequate if we have, say, two
macroion surfaces with regularly placed charges of $-1$, and there
are ions of charge $+2$ between them. However appealing, the
bridge concept becomes increasingly fuzzy when $Z$-ions have
charges of $3$ or higher and when charges in the macroion are not
positioned regularly.

\begin{figure}
\epsfxsize=5.5cm \centerline{\epsfbox{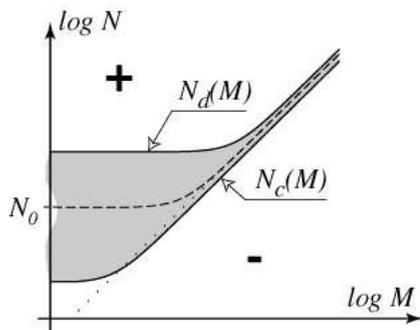}} \caption{Generic
phase diagram of reentrant condensation and charge inversion in
terms of macroion concentration $M$ and $Z$-ion concentration $N$
\protect\cite{Shklov005}.  Isoelectric composition is shown by the
dotted line.  The dashed ``neutrality line'' corresponds to
neutral complexes in the dilute phase. The segregation region is
shaded. Minus and plus indicate the signs of complexes of DNA with
$Z$-ions.} \label{fig:reentrant}
\end{figure}

Motivated by experimental observations of DNA condensation (see
Sec. \ref{sec:experiment}), there was a significant effort by
theorists to try to explain attractive forces by going beyond the
bridge model. For simplicity, and following the majority of works,
let us consider two planar macroion surfaces with some $Z$-ions
between them.  Of course, Poisson-Boltzmann theory predicts pure
repulsion for such a system.   However, attraction was observed in
several computer experiments, including Guldbrand et al (1984),
Kjellander and Marcelja (1985), Gronbech-Jensen et al (1997),
Linse and Lobaskin (1998), and Moreira and Netz (2000b). On the
theory side, several groups attempted to go beyond the mean field
approximation.   An important observation is that due to dynamic
fluctuations of counterions, there is an attractive component
(similar to Van der Waals interactions), but at effectively high
temperature, or small $\Gamma$ (see Eq. for the definition of
(\ref{eq:Gamma})), the Poisson-Boltzmann repulsion still dominates
and the force remains mainly repulsive (Oosawa, 1968; Lau and
Pincus, 1998; Ha and Liu, 1998; Podgornik and Parsegian, 1998;
Golestanian et al, 1999; Golestanian and Kardar, 1999).

On the other hand, attraction of like charges dominates at
effectively low temperatures, when $\Gamma \gg 1$, and the idea of
spatial correlations between $Z$-ions, which is the central idea
of this Colloquium, sheds light on the nature of this attraction.
Indeed, for extremely large $\Gamma$, we deal with two Wigner
crystals on the two opposing plates; they gain energy when
approach each other by properly positioning themselves in the
lateral direction. This was shown by Rouzina and Bloomfield (1996)
(see also Gronbech-Jensen et al, 1997; Levin et al, 1999; Moreira
and Netz, 2000b). Furthermore, Gronbech-Jensen et al (1997) and
Shklovskii (1999a) pointed out that the long range order of a
Wigner crystal is not important for this attractive force. As in
the case of charge inversion, what is important is correlation and
short range order. As we know, $Z$-ions form a SCL on the macroion
surface as soon as $\Gamma$ becomes large.  Imagine now bringing
two planar surfaces, along with their respective SCL, very close
to each other.  What we have now is essentially two copies of Fig.
\ref{fig:WC} superimposed on top of one another, with $Z$-ions
confined in between. Clearly, two SCL merge, lowering the energy
per $Z$-ion from $\varepsilon (n)$ to $\varepsilon ( 2 n) <
\varepsilon (n)$
(see Eq. (\ref{benergy})).  
Physically, every $Z$-ion in the merged SCL is sandwiched between
two macroion surfaces, and its Wigner-Seitz cell can be
approximated as a pair of discs, one on each surface (see again
Fig \ref{fig:WC}).  The charge of the cell must still be $-Ze$,
but since there are two surfaces, the radius of the cell is
reduced by the factor $1/\sqrt{2}$, leading to the energy gain. In
some sense, this theory returns us to the idea of bridges, albeit
on a completely new level, with each $Z$-ion bridging between two
sides of its Wigner-Seitz cell, which can include many surface
charges.

These arguments hold, at least qualitatively,  not only for
plates, but also for macroions of other shapes, including DNA
double helices.  To be specific, consider two DNA double helices.
When the concentration of $Z$-ions is equal to $N_0$, each DNA is
neutralized by $Z$-ions, and the two neutral complexes attract
each other at short distances.  What happens if the concentration
of $Z$-ions is higher or lower than $N_0$?   In this case,
correlation-induced attraction, which is short-ranged, competes
with Coulomb repulsion, which is much longer-ranged. Note that the
Coulomb repulsion force is present both at $N < N_0$, when the
DNAs are partially screened by $Z$-ions and negative, and at $N >
N_0$, when they are overcharged and positive.

What are the implications of this competition between attraction
and repulsion?  They are summarized in Fig. \ref{fig:reentrant}
which shows a phase diagram of the solution with number
concentrations of macroions, $M$, and $Z$-ions, $N$ (along with
the neutralizing amount of monovalent ions and salt.)  The major
feature of the phase diagram is the segregation region, which is
the shaded area in the Fig. \ref{fig:reentrant}.  As the figure
indicates, the generic scenario is that of {\em reentrant
condensation}.  DNA molecules stay in solution and remain negative
at $N < N_c(M)$, forming undercharged complexes with $Z$-ions. At
some concentration of $Z$-ions, $N = N_c (M)$, repulsion loses to
the correlation attraction and a condensed phase of DNA is formed,
coexisting with a dilute phase. The condensed phase for DNA
represents a (nematic) bundle of helices, it exists in the
interval $N_c(M) < N < N_d(M)$.  Finally, at $N = N_d(M)$,
repulsion overcomes the correlation attraction, and the DNA
molecules dissolve and form positive (overcharged) complexes with
$Z$-ions.

Inside the coexistence region, there is a neutrality line, on
which the equilibrium dilute phase consists of {\em neutral}
complexes.  At very small DNA concentrations, the neutrality
condition corresponds to the concentration $N_0$ of $Z$-ions (Eq.
\ref{eq:N0}).  To see what happens at larger DNA concentration,
consider increasing $M$, starting from the overcharged complexes,
well above segregation on the phase diagram in Fig.
\ref{fig:reentrant}. When $M$ grows, the solution runs out of
$Z$-ions when it approaches the ``isoelectric line'' $- \eta L M +
ZeN = 0$, with $- \eta < 0$ and $L$ being the DNA linear charge
density and length, respectively.  Near this line, the charge of
complexes flips sign.  Thus, the neutrality line crosses over from
$N = N_0$ to the isoelectric line.  The border lines $N_c(M)$ and
$N_d(M)$ follow a similar pattern.  Although not plotted in Fig.
\ref{fig:reentrant}, at extremely small values of $M$ these two
lines join together at a critical point, and for smaller $M$ only
intramolecular condensation of DNA (the coil-globule transition)
is possible if the DNA molecule is long enough.

We have considered the phase diagram in Fig. \ref{fig:reentrant}
for a solution of DNA chains with small $Z$-ions. As a matter of
fact, the diagram is qualitatively quite general \cite{Shklov005}.
For instance, it applies to a solution of DNA with large
positively charged particles.  In Sec. \ref{sec:flexible}, we
considered the case of a small DNA concentration, $M$, and a large
concentration of spheres, $N$, which corresponds to the region
above the coexistence region on the phase diagram in Fig.
\ref{fig:reentrant}. We found that complexes have the form of
necklaces, as shown in Fig. \ref{fig:necklace}, and that they are
overcharged, i.e., contain more spheres than necessary to
neutralize DNA molecule. Large spheres are so strongly bound to
DNA that the concentration $N_0$ for them is extremely small and
any real experiment deals with the narrow upper-right part of the
diagram. Suppose now that there are relatively few spheres in the
solution, so we are below the neutrality line. In this situation,
chains make an overcharging number of turns around each sphere.
This is energetically favorable due to the repulsive correlations
between subsequent turns on a sphere surface.  The inverted net
charge of each sphere is about as large as for the case of a
single sphere discussed in Sec. \ref{sec:flexible}. Furthermore,
the inverted charge of spheres determines their distribution along
the chain of polyelectrolyte. Negative spheres repel each other
and, therefore, the complex once again has the periodic
beads-on-a-string structure, Fig. \ref{fig:necklace}, which
resembles the 10 nm chromatin fiber.  In the narrow vicinity of
the neutrality line, even a small correlation attraction between
touching spheres is sufficient to drive aggregation (or
coil-globule collapse for long chains) of DNA with spheres.

\section{Experimental evidence of charge
inversion}\label{sec:experiment}

How does the theory of correlated screening compare with
experiment?  For the purposes of this Colloquium, we restrict
ourselves to a qualitative comparison only.

\begin{figure}
\epsfxsize=7.5cm \centerline{\epsfbox{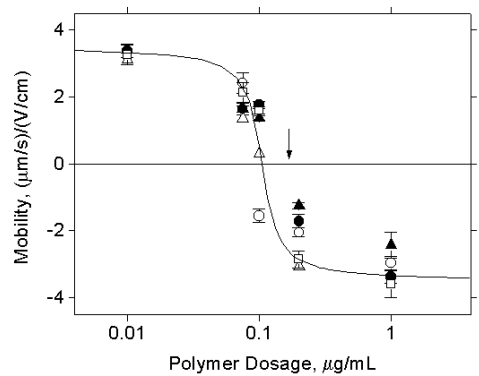}}
\caption{Mobility of positive latex particles (macroions) in the
presence of $0.005$ M (moles per liter) concentration of NaCl as a
function of polymer mass concentration.  Polymers (single-stranded
DNA chains) play the role of $Z$-ions.  Different symbols
correspond to DNA of the following lengths (in monomers):
$\blacktriangle$ - 8, $\circ$ - 10, $\triangle$ - 40, $\bullet$ -
80, $\square$ - 1400.  The line is drawn to guide the eye.  The
arrow indicates the isoelectric point, the polymer mass
concentration of $0.17$ $\mu$g ml$^{-1}$ at which DNA charge
neutralizes the latex particles.} \label{fig:GrantData}
\end{figure}

First observations of charge inversion were reported a long time
ago by De Jong (1949).  He was able to measure electrophoresis of
macroscopic aggregates in the phase segregation region of the
phase diagram Fig \ref{fig:reentrant}, and observed the reverse of
their mobility  upon crossing the isoelectric line.  More
recently, Kabanov and his co-workers (Kabanov and Kabanov, 1995;
Kabanov et al, 1996; Kabanov and Kabanov, 1998) examined mixtures
of positive and negative polymers and directly observed the
interpolyelectrolyte complexes in which a larger polymer of one
sign (playing the role of a macroion) was seen to bind an
overcharging amount of smaller polymers of the opposite sign
(playing the role of $Z$-ions).  The effect was directly seen due
to the reversal of the electrophoretic mobility of complexes.
Furthermore, Wang et al (1999) observed similar reversal for a
mixture of polyelectrolytes (macroions) and micelles ($Z$-ions).
Gotting et al (1999) found the reversed mobility for the
nanoparticles (macroions) and short single stranded DNA (elongated
$Z$-ions). Walker and Grant (1996) demonstrated this phenomenon
for 120 nm latex particles (macroions) with single stranded DNA
($Z$-ions) ranging from 8 to 1400 nucleotides; their data are
presented in Fig. \ref{fig:GrantData}.  Kabanov et al (2000) and
Evans et al (2001) observed the reversed electrophoretic mobility
for DNA with the dendrimers as $Z$-ions.

An interesting observation, apparent from Fig.
\ref{fig:GrantData}, is that the data for different DNA lengths
collapse onto a single master curve upon re-scaling (in which
mobility is plotted against mass concentration of DNA instead of
number concentration, $N$).  We note in passing that this
observation can be rationalized by the following argument.
Isoelectric point is obviously determined by the number of charged
groups per unit volume, but this quantity is insensitive to the
overall length of DNA and is simply proportional to the mass
concentration.

All of the above mentioned experimental works rely on the reverse
of electrophoretic mobility as an indication of charge inversion.
Indeed, this is conceptually the most straightforward approach. It
is valid because $Z$-ions are strongly bound to the macroion, with
energies larger than $k_B T$, and move together with it.  On the
other hand, monovalent ions screening the net charge $Q^*$ are
attracted to macroion with energy much smaller than $k_B T$ and
move in electric field in the opposite direction.  For these
reasons, it is the net charge $Q^{*}$ that determines both
magnitude and sign of the observed electrophoretic mobility. This
remains correct also in the case when monovalent ions adsorb on
$Z$-ions, effectively reducing them to $Z^{\prime}$-ions with
$Z^{\prime} < Z$, as in Fig. \ref{fig:Motohiko}.  In all cases,
the net charge includes all ions bound with energies in excess of
$k_B T$.  In a recent molecular dynamics simulation, Tanaka and
Grosberg (2001b) have directly examined the mobility of charge
inverted macroion complexes similar to the one shown in Fig.
\ref{fig:Motohiko}.  They confirmed that adsorbed $Z$-ions drift
together with the macroion in a weak electric field and that the
sign of the net charge $Q^{*}$ determines the direction of
electrophoresis.

It is also worth saying that the interpretation of electrophoretic
experiments on charge inversion is not affected by the recent
discoveries of Long et al (1996 and 1998).  These authors noted
that electrophoretic mobility, under some circumstances, may not
be entirely determined by the charge.   For instance, even an
overall neutral object may move in an electric field provided that
it has strong asymmetry of charge distribution.  A simplest
example consists of two balls of different radiuses, rigidly
connected by a thin rod, and having opposite charges. The effect
is due to the fact that external electric field acts not only on
macroion itself, but also on surrounding co- and counterions,
causing the latter to flow and to exert viscous friction drug
forces on the macroion.  Strong geometrical asymmetry of positive
and negative charges leaves these drug forces unbalanced.  In this
colloquium, we discuss only macroions in which the bare charge is
uniformly distributed on the surface. Then, the charge of $Z$-ions
is practically uniform, too (apart from the small length scales of
the order of $R$, the distance between neighboring adsorbed
$Z$-ions).  For such macroions, reversal of electrophoretic
mobility does indeed indicate the inversion of charge.

For a more detailed comparison with experiments, we should
remember that charge reversal is expected to be accompanied by
coagulation, as discussed above in Sec. \ref{sec:forces}. Whether
in equilibrium or not quite in equilibrium, these large complexes
should scatter light strongly. There are many experiments
reporting such observations.

Let us begin with DNA. It has been known for some time that at
some critical concentration of $Z$-ions, $N_c$, DNA abruptly
condenses into large bundles \cite{Bloomfield}. Recently it was
discovered that at a much larger critical concentration, $N_d$,
the bundles dissolve back (Saminathan et al, 1999; Pelta et al,
1996a; Pelta et al, 1996b; Raspaud et al, 1998; Raspaud et al,
1999). Specifically, for the spermine ions ($Z =4$), it was found
experimentally that $N_c = 0.025$~mM and $N_d = 150$~mM.  If one
interprets these data in the framework of the theory of Nguyen et
al (2000c), the experimental values of $N_c$ and $N_d$ imply that,
for spermine, $N_0 =3.2$~mM and the binding energy of two helices
per one spermine ion is $u = 0.3 k_B T$. The last value agrees
with the one obtained by a different method \cite{Rau}.

Let us now discuss some other systems. Wang et al (1999) studied
complex formation in mixture of micelles and oppositely-charged
polyelectrolytes.  In this experiment, the total charge of
micelles was controlled by changing the concentration of the
cationic lipid in the solution.  In agreement with the above
theory, measurements of dynamic light scattering and turbidity
(coefficient of light scattering) show that complexes condense in
bundles and the solution coacervates in the vicinity of the point
where mobility crosses over between two almost constant values,
positive and negative.

For the complexes of latex particles with DNA of various lengths,
examined by Walker and Grant (1996), equilibrium conditions were
not found, but a significant rate of aggregation of latex
particles was observed in the same narrow range of DNA
concentrations where the mobility flips the sign.

There is a large body of interesting experimental (R\"adler et al,
1997; Koltover et al, 1999) and theoretical (Harries et al, 1998;
Bruinsma, 1998) work on phase diagrams and overcharging of
lamellar cationic lipid-DNA self-assembled complexes.  These
solutions always seem to have aggregates due to hydrophobic
attraction of lipids.  A phase diagram of this kind has been
sketched by R\"adler (2000).

Another large group of works with nice (and technologically
useful) examples of overcharging deals with the sequential
adsorption of multiple layers of polyelectrolytes of alternating
sign (see Decher (1997) and references therein).  Say, a positive
surface is first treated with negative polymers, gets overcharged
and becomes negative, then it is treated with positive polymers,
gets overcharged etc. This procedure works reliably with up to
many tens of layers.  Theoretical interpretation of this technique
was discussed by Castelnovo and Joanny (2000).  It falls beyond
the framework of the present Colloquium as it involves certain
kinetic considerations while we deal in this Colloquium only with
equilibrium phenomena.

\section{Correlations "in a sheep's skin"}\label{sec:other_mechanisms}

We mentioned correlations so many times in this Colloquium, that
the reader may want to ask: Are there alternative,
correlation-independent, electrostatic mechanisms leading to
charge inversion?  Our answer is no, and we argue that the
correlation-based mechanism is the universal one.  This fact
notwithstanding, we should say that correlations may show up in a
number of ways, sometimes masked like a wolf in a sheep's skin.

To understand better the role of correlations, let us first
consider the case of no correlations.  Namely, suppose, we have a
set of randomly positioned point-like charges, equal numbers of
$+e$ and $-e$.  It is easy to establish that the averaged
interaction energy in such system is exactly zero, where average
is taken over independent random positions of all charges.
Similarly, averaged electric field in the system is also zero.
Both these statements remain correct also for the one component
plasma, in which point-like charges of one sign are randomly
positioned on the smeared uniform background of the opposite sign.

The essence of screening is that charges in plasma are {\em not}
positioned randomly.  Correlations happen because ions reconfigure
themselves non-randomly to gain some energy.  This so-called
correlation energy is well known in plasma physics \cite{LL}. It
is negative, meaning that correlated configuration is lower in
energy, or more thermodynamically favorable, than random
configuration.

These simple facts allow us to understand the underlying role of
correlations in one of the theories suggested in the literature to
explain charge inversion.  This theory, put forward by Park et al
(1999), views monovalent {\em counterion release} from $Z$-ions as
the driving force behind charge inversion. We argue that while
counterion release accompanies, it is itself driven by
correlations. As we did already more than once, let us imagine
that DNA molecules, along with their Onsager-Manning-condensed
small ions, are being adsorbed on the macroion one at a time
(possibly releasing some of their counterions).  Let us further
consider the moment when the neutralization condition has just
been achieved. Pretend now that DNA rods ($Z$-ions) are
distributed randomly on the surface, uncorrelated in both
positions and orientations.  In this case, next arriving DNA
molecule feels no average field, so that it has no reason to
release its counterions. The situation is completely different if
DNA molecules are correlated on the surface (see Figs.
\ref{fig:dna2} or \ref{fig:rods}), where locally each molecule is
surrounded by a correlation hole - the positive stripe of the
background charge (the Wigner-Seitz cell). The corresponding
field, or the positive potential of the Wigner-Seitz cell, causes
the release of counterions from DNA not only at the neutrality
point, but even if the surface overall is overcharged (see the
solution of this problem given by Eqs. (\ref{NL1}), (\ref{NL2})).
These qualitative arguments can be formulated also more
quantitatively \cite{Nato}.  Thus, correlation hole, or adjustment
of DNA molecules to each other, or image charge, or correlations
(all synonyms!) is the driving force for {\em both} counterion
release {\em and} charge inversion; under a sheep's skin of
counterions release, there is a wolf's face of correlations.

Let us now discuss another approach which we call {\em
metallization}. It was pioneered by Mateescu et al (1999), who
considered complexation of a polyelectrolyte with a sphere, and by
Joanny (1999), who examined adsorption of flexible polymers on a
charged plane.  Metallization theory considers adsorbed $Z$-ions
as a continuous medium similar to a metal, while still treating
the bulk solution as consisting of discrete charges.

We argue that metallization theory is in fact an overestimate of
correlation effects.  Indeed, as we illustrated with Fig.
\ref{fig:coming_ion}, SCL behaves as a metal on the length scale
above $R$, the distance between neighboring adsorbed $Z$-ions.  It
behaves as a metal in the sense that it responds to the
approaching new ion by forming an image.  Clearly, the smeared
continuum is also a metal in this sense, and even a better metal -
it is a metal on all length scales.  Another way to view that same
physics is to note that correlation suppresses electric field of
every $Z$-ion beyond certain distance of order $R$, while for the
smeared continuum the field is suppressed everywhere and,
therefore, the effect is overestimated.

The latter view suggests also another fruitful interpretation of
metallization approach, which is in terms of the self-energy of
$Z$-ions.  Here, we resort to the terminology in which self-energy
of an ion is identified with the energy of the electric field
outside certain cutoff length, such as an ion size. (We are not
interested in the electric field on the smaller length scales:
although energy of this field is infinite, it does not change upon
any processes considered here, such as adsorption of $Z$-ions.)
Using this language, the metallization theory becomes physically
transparent. Indeed, in this theory adsorption of $Z$-ions is
energetically favorable because they do have self-energy while in
the bulk solution and loose it completely upon adsorption.  Once
again, real screening, or correlations between $Z$-ions,
correspond to suppression of a part of self-energy, corresponding
to the energy of the electric field beyond the distance of order
$R$. Energy of the field between the ion size and $R$ is the
amount of overestimation by the metallization theory.  By
contrast, the Poisson-Boltzmann approximation fails to describe
charge inversion precisely because it smears $Z$-ions everywhere
and neglects their self-energies.

Representation of correlations in terms of self-energy allows us
to address one more practically important problem, namely, the
discreteness of charges (Nguyen and Shklovskii, 2001e).  Indeed,
instead of uniformly charged surface, like Fig.
\ref{fig:WCpicture}, it would be closer to reality (and to a
chemist's heart) to think of a macroion as having some charged
groups with unitary charge each.  For instance, negative ($-e$)
charges of DNA are positioned on an external spiral rim of the
double helix, at the distance $A = 0.67$ nm from each other along
the rim.

\begin{figure}
\epsfxsize=8cm \centerline{\epsfbox{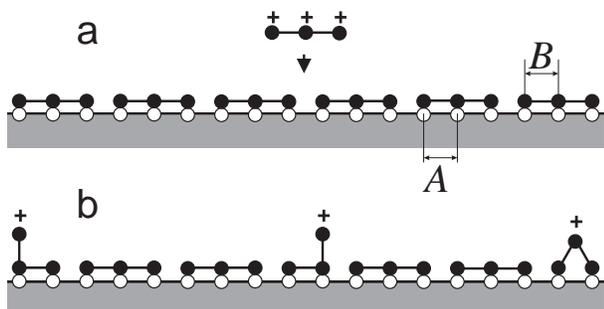}}
\caption{Schematic representation of charge fractionalization. a)
One strand of negative charges of DNA (empty balls) is completely
neutralized by positive $Z$-ions with $Z=3$, their charged groups
are shown by black balls. A new $Z$-ion is approaching.  b) The
new $Z$-ion is "digested."  Its charge is split in $+e$ charges of
$Z$ defects, tails and archs. } \label{fig:fractionalization}
\end{figure}

To be specific, let us consider a macroion which is a regular
lattice of charges $-e$ ("unfolded DNA strand") and $Z$-ion having
also a linear array of charges $+e$ (a short polyelectrolyte).
Importantly, $Z$-ions always have some degree of flexibility; for
instance, in the case $Z = 3$, as shown in Fig.
\ref{fig:fractionalization}, we can imagine that they freely bend
in the middle.  For simplicity we assume that the distance between
charges in the $Z$-ion, $B$, matches exactly that in the macroion,
$A$: $A = B$. Finally, we assume that the $Z$-ion charges and the
macroion charges can approach each other to the minimal distance
much smaller than $A$. Then $Z$-ions can attach to the macroion
locally compensating each charge and, therefore, achieving
complete neutralization, as we show in Fig.
\ref{fig:fractionalization}a.  The neutralization is so perfect
that it is difficult to imagine how another $Z$-ion can be
attached.  Fig. \ref{fig:fractionalization} explains why this
happens. Similarly to Fig. \ref{fig:coming_ion}, suppose that the
macroion is already neutralized by $Z$-ions and a new $Z$-ion
comes.   Then it turns out energetically favorable to disturb
order among the neutralizing $Z$-ions by $Z$ defects in $Z$
independent places, thus opening a room for a new $Z$-ion.  In
each defect, one charge of $Z$-ion is detached from the
corresponding macroion charge, forming positive tail or arch above
the surface and leaving negative vacancy on the macroion. Then,
shifting $Z$-ions along the macroion, $Z$ vacancies can join
together and form a large vacancy capable to accommodate an entire
new $Z$-ion. A net result is that $Z$ disconnected charges $+e$
appear on top of completely neutralized macroion (Fig.
\ref{fig:fractionalization}b), or, in other words, the charge of
$Z$-ion is fractionalized. To avoid misunderstanding, we emphasize
that none of the chemical bonds is really cut.

Fractionalization effectively eliminates self-energy of the free
$Z$-ion. Indeed, self-energy of the $Z$-ion is simply the energy
of repulsion between $Z$ positive constituent charged groups in
the extended conformation which the $Z$-ion assumes in the
solution. In the fractionalized state, charges are far apart and
practically do not interact, so that self-energy is gained.  These
results are easily generalized for the case $B < A$, when $Z$-ions
have larger linear charge density than macroion.  For instance, if
$B=A/2$, then $Z$-ions repelling each other form a "Wigner
crystal" on top of the lattice of macroion charges where they
alternate with vacant places (similarly to Figs.
\ref{fig:WCpicture}, \ref{fig:WC}, \ref{fig:coming_ion}, and
\ref{fig:rods}).

Is fractionalization a correlation-independent mechanism of charge
inversion? Of course, not: this phenomenon is solely due to the
correlated distribution of $Z$-ions, which avoid each other at the
macroion. Fractionalization is yet another mask under which
correlations may show up.

\section{Charge inversion in a broader physics
context}\label{sec:analogies}

In conclusion, we would like to show that charge inversion studied
in this Colloquium has many physical analogies. There are other
``charge inverted'' systems in physics.  Let us start from the
hydrogen atom. It is known that it can bind a second electron,
forming the negative ion H$^{-}$ with an ionization energy of
approximately $0.05Ry$ \cite{BookAboutNegativeIons} . We can
consider this effect as the inversion of proton charge. Attraction
of the second electron to the neutral atom is due to the Coulomb
correlation between electrons: the first electron avoids the
second one, spending more time on the opposite side of the proton.
In other terms one can say that binding is related to polarization
of the neutral core.

Negative ions - nuclei overcharged by electrons - exist also for
larger atoms.  Mean field Thomas-Fermi or Hartree theories fail to
explain negative ions \cite{LLQM}. One must include exchange and
Coulomb correlation holes to arrive at a satisfactory theory
explaining bound state and the nonzero ionization energy of a
negative ion \cite{BookAboutNegativeIons}.  The Thomas-Fermi
theory of an atom is an analog of the Poisson-Boltzmann theory of
electrolytes. It is not surprising that both fail to explain
charge inverted states.

Similar considerations apply also for a macroscopic metallic
particle. Electrons in such a particle have a negative (compare to
vacuum) chemical potential or, in other words, positive work
function. The work function is known to vanish in the Thomas-Fermi
or Hartree approximations \cite{Lang}.

The energy $|\mu_{WC}|$ plays the same role for $Z$-ions on the
insulating macroion surface as the ionization energy of a negative
ion or the work function of metallic particle for electrons.
Similarly to electrons, charge inversion of a charged insulating
macroion by $Z$-ions can not be obtained in the mean field
Poisson-Boltzmann approximation. Only correlations of $Z$-ions on
the surface of the macroion can lead to charge inversion.

Let us now return to the Onsager-Manning condensation (Manning,
1969; Sec. \ref{sec:history}). Kosterlitz and Thouless (1972)
discovered a similar threshold phenomenon for generation of free
vortexes in two dimensional superfluids or superconductors. They
noticed that due to the logarithmic form of attractive
interaction, two vortexes of opposite sign decouple only above
some critical temperature, $T_{KT}$.  Later, Kosterlitz-Thouless
theory was applied to unbinding of dislocations and disclinations
in the theory of defect-induced melting of two-dimensional
crystals (Nelson and Halperin, 1979; Young, 1979).

In the Kosterlitz-Thouless theory one can identify analog of the
short-range correlation contribution to the chemical potential of
$Z$-ions (which we called $|\mu_{WC}|$).  This is the energy of
creation of the two vortex cores. Similarly to $|\mu_{WC}|$,  this
energy provides additional binding of vortexes and strongly
reduces the concentration of free vortexes at $T > T_{KT}$
\cite{Minn}. In contrast to the Kosterlitz-Thouless theory, the
short range contribution $|\mu_{WC}|$ was only recently introduced
\cite{Perel99}.

Another physical analogy already mentioned in Sec.
\ref{sec:layers} is the integer quantum Hall effect.  We should
add that charge fractionalization, as illustrated by Fig.
\ref{fig:fractionalization}, is an analog of the fractional
quantum Hall effect \cite{frac}.  Finally, Fig.
\ref{fig:fractionalization} bears also analogy with electron
charge fractionalization in the polyacetilene \cite{Braz}.

\section{Conclusions and outlook}\label{sec:conclusion}

To conclude, we have discussed the physical picture of screening
for the case of strongly interacting ions.  This case appears to
have been overlooked for many decades, since Debye.  Its
theoretical study, motivated by the experiments mainly on gene
delivery and chromatin structure, revealed new interesting
physical insights. Specifically, correlations between screening
ions lead to such counterintuitive phenomena as charge inversion,
reverse electrophoretic mobility, attraction of like charged
molecules or colloids, etc.  The physical theory of these
phenomena is aesthetically attractive as it presents many
parallels with other areas of physics, ranging from quantum Hall
effect to atomic physics and metals. The potential applications
are many, in both chemical and biological realms. We can mention
here all sorts of manipulations with DNA, both for biological
purposes and for using DNA as an assembly tool for non-biological
nanotechnology.  They all require understanding of electrostatic
properties of DNA chains.  Many diseases have to do with
mis-assemblies of charged proteins, such as actin; we need to
understand better the assembly  of such objects.  Food, cosmetic,
paper, and waste water treatment industries, are all about charged
colloids, and the list of applications is easy to continue. In
brief, this theory is one of the busy junctions where physics
meets chemistry and biology.

\acknowledgements

We enjoyed collaboration with V. I. Perel, I. Rouzina, and M.
Tanaka.  We are grateful to V. Bloomfield, E. Braun, R. Bruinsma,
P. Chaikin, A. Dobrynin, P. Dubin, M. Dyakonov, W. Gelbart, S.
Girvin, W. Halley, C. Holm, J.-F. Joanny, A. Kabanov, V. Kabanov,
A. Khokhlov, R. Kjellander, K. Kremer, A. Koulakov, V. Lobaskin,
F. Livolant, D. Long, G. Manning, R. Netz, P. Pincus, R.
Podgornik, E. Raspaud, M. Rubinstein, J.-L. Sikorav, U. Sivan, M.
Voloshin, and J. Widom for useful discussions. We thank Zhifeng
Shao for the permission to use Figs. \ref{fig:dna2} and
\ref{fig:chromatin1} and S. Grant for the permission to use Fig.
\ref{fig:GrantData}. T. T. N. and B. I. S. are supported by NSF
DMR-9985785.


\end{document}